\documentclass[12pt]{article}

\usepackage{amssymb,amsmath,amsthm,amsfonts}

\usepackage{array,longtable}
\usepackage[dvips]{graphicx}
\usepackage[dvips]{color}
\usepackage{epsfig}
\usepackage[mathscr]{eucal}
\def\beq{\begin{equation}}                     %
\def\eeq{\end{equation}}                       %
\def\bea{\begin{eqnarray}}                     
\def\eea{\end{eqnarray}}                       
\def\PL{{\it Phys. Lett.} }                    
\def\PR{{\it Phys. Rev.} }
\def\PRL{{\it Phys. Rev. Lett.} }
\def\APJ{{\it Astroph.~J.~}}
\def\AJ{{\it Astron.~J.~}}                 %

\newcommand{\pd}{\partial}

\def\+{^\dagger}

\def\+{^\dagger}

\newcommand{\norm}[1]{\raise.3ex\hbox{:}#1\raise.3ex\hbox{:}}

\begin {document}
\title{
\Large{ \textup{Nonlocal  String Tachyon as a Model for
Cosmological Dark Energy}}}
\author{
I.~Ya.~Aref'eva
\vspace{5mm}\\
Steklov Mathematical Institute\\
Russian Academy of Sciences\\
Gubkin st. 8, Moscow, 119991, Russia \ } \maketitle

\begin{abstract}
There are many different phenomenological models describing the
 cosmological dark energy and accelerating Universe by choosing adjustable
functions. In this paper we consider a specific model of scalar
tachyon field which is derived from the NSR string field theory
and study its cosmological applications. We find that in the
effective field theory approximation the equation of state
parameter $w<-1$, i.e. one has a phantom Universe. It is shown
that due to nonlocal effects there is no quantum instability that
the usual phantom models suffer from. Moreover due to a flip
effect of the potential the Universe does not enter to a future
singularity.

\end{abstract}

\newpage
\section{Introduction}

It was suggested by Ia Supernova observations
that the Universe is presently  accelerating\cite{Perlm,Riess}. The basic sets of experiments
now includes also the
Cosmic Microwave Background (CMB) anisotropies,  X-ray data from galaxy clusters,
large scale structure and age estimates of global clusters, for a review  see
\cite{Spergel,S-St,Sachni}.
It is believed that a new particle and/or gravitational physics is required to
explain the acceleration of
the expansion of the Universe.
The observations suggest that the bulk of energy density in the Universe is
gravitationally repulsive and appears
like an unknown form of energy (dark energy) with negative pressure. It is believed that
2/3 of the total density of the universe is in a form of dark energy.

There exist many different models of dark energy. It is convenient
to describe them by using the equation of state parameter
$w=p/\rho$, where $p$ is a pressure and $\rho$ is the energy
density. The analysis of the current observation data  shows that
$w$  lies in the range $-1.61<w<-0.78$
\cite{knop,Spergel,Tegmark}.

A list of dark energy models includes (see \cite{S-St}-\cite{kitaj} and refs. therein)
\begin{itemize}
\item $w=-1$: the cosmological constant;
\item $w=const \neq -1$: the cosmic strings, domain walls, etc.;
\item $w \neq const$: quintessence scalar field, chaplygin gas, k-essence, Dirac-Born-Infeld(DBI)
 action,
braneworlds, etc.;
\item $w < -1$: phantom models.
\end{itemize}

The most challenging for theoretical physics would be the case
$w < -1$ \cite{Caldwell}-\cite{Melchiorri} (see \cite{Frampton,Sachni,Padmanabhan-rev}
for  reviews).
In this case the weak energy condition ($\rho>0, \rho+p>0$)
is violated and some strange phenomena as negative entropy and temperature appear.
In many models the phantom
Universe in finite time ends up in the singularity called the Big Rip \cite{Caldwell03}).

 In this paper we consider string field theories
nonlocal tachyon where the condition $w < -1$ is realized.
We show that due to a peculiar properties of
nonlocal tachyon dynamics we can get a phantom
universe without problems with unstability.
A nonlocal dynamics for the tachyon field is obtained by truncation
of the covariant string field equations. We study a nonlocal
dynamics of a string tachyon in the cosmological Friedmann metric. The string
theory that we have in mind is the NSR string theory compactified
on a six dimensional compact manifold. Moreover we assume that all
moduli are frozen and unstable non-BPS brane extends along the
three large spatial dimensions. The tension of the 3-dimensional
non-BPS brane acts as the 4-dimensional universe cosmological
constant and the tachyon dynamics describes a deviation from a pure cosmological
constant regime.

A  presence of  tachyons was considered
 as a main drawback of corresponding  strings theory. Bosonic string has
a tachyon and its absence in superstrings
 was a main motivation to introduce superstrings. However few years ego tachyon
has  found an application in the context of brane scenarios. The
open bosonic string and GSO $-$ NSR string tachyons were used to
describe D-brane decays. According to the Sen conjecture in the
perturbative vacuum there are instable branes
filling space-time and all these branes disappear in the true vacuum (for
review see \cite{review-sft},\cite{Sen-g}).
Rolling tachyon solutions
describe transitions from perturbative vacuum to non-perturbative one.

Here we use the covariant
 open string field
theory(SFT) approach \cite{review-sft}  to tachyon dynamics (about
the effective  DBI approach to tachyon dynamics
see \cite{Sen},\cite{0301117},\cite{Sen-g} and refs. therein).
A crucial feature of  the rolling tachyon
solution for non-BPS brane \cite{AJK} is that it interpolates
between two non-perturbative vacua with the same energy. We argue
that at large time  the dynamics is governed   by an effective
action that contains a ghost kinetic term, in spite of absence of
ghosts in the nonlocal action. We emphasize that the ghost here is
a result of an approximation to the exact nonlocal action. There
is no ghost in the nonlocal action. So we get a phantom universe
as an approximation to the true nonlocal dynamics.
This is the
reason why there are no pathologies in this scenario with the thermodynamical
instabilities.

The rolling
tachyon solution passes  the perturbative vacuum with non-zero
velocity. Therefore, to reach the true vacuum starting from the
perturbative one we have to
supply  the tachyon with a large enough initial velocity .
This is a rather non-standard situation from the
local field theory point of view, where
one does not have to push strongly the  tachyon to make it reach the true vacuum,
it is need just an infinitesimal small perturbation.
This effect occurs due to the presence of derivative terms in the
interaction. These terms may be studied in an effective action
approximation, which corresponds to keeping only few
terms of an expansion of an nonlocal operator.
We note  that to reach the nonperturbative vacuum
one has to add to the action a brane tension which is
larger that is required by the Sen hypothesis.
This large  brane tension can be interpreted as an
effect of the closed string excitations.

We find that in the effective field theory
approximation the equation of state parameter $w<-1$, i.e. one has
a phantom Universe.
In the DBI approach  the equation
of state parameter interpolates
 between -1 at early time and 0 at a later time.
 Within application of the DBI action to cosmology
 there are  problems with large density perturbations,
 reheating and caustics formation \cite{Kofman,FKS}.

 The paper is organizes as follows.
In Sect.  2.1. the nonlocal tachyon action is written in
the Friedmann background.
In Sect. 2.2 we shortly remind a construction of the rolling tachyon solution
  for non-BPS brane \cite{AJK}.
In Sect. 2.3 we study the tachyon dynamics in the Friedmann metric in the
effective action approximation.
In this approximation one gets $w<-1$.
In Sect. 2.4 we show that to reach a nonperturbative vacuum
one has to add to the action a brane tension
which is larger that is
required by the Sen hypothesis.
In Sect. 3 we show the stability of the
nonlocal tachyon model in the true vacuum.

\section{Non-BPS tachyon in Friedmann space-time}
\subsection{General set up}
We consider a non-BPS tachyon leaving on 3-brane and interacting with gravity
with the following action
\begin{equation}
\label{action-E-M} S = \frac{M_p^2}{2}\int \sqrt{-g}d^4x\,
R+S_{tach}  \ .
\end{equation}
where
\begin{equation}
\label{act-met}
S_{tach}= \int\; \sqrt{-g}d^{4}x
\left(-\frac{q^2}{2} g^{\mu
\nu}\pd_\mu \phi \pd_\nu \phi +\frac12 \phi^2-\frac14\Phi^4
\right),
\end{equation}
$\Phi=\exp(\frac12\Box_g) \phi$,
$\Box_g=\frac{1}{\sqrt{-g}}\pd _\mu \sqrt{-g}g^{\mu\nu} \pd _\nu
$, $q^2=const<1$.
Here we assume that  all constants are absorbed into
$M^2_p$. The action (\ref{act-met}) generalizes the non-BPS tachyon action
obtained from low level truncated SFT to the case of a non-flat metric \cite{AJK}.

On  space homogeneous configurations  in the Friedmann metric
\beq
\label{met}
ds^2=-dt^2+a^2(t)(dr^2+r^2(d\theta^2+\sin^2 \theta d\phi ^2))
 \eeq
 the action (\ref{act-met}) takes the form
 \begin{equation}
S_{tach}[\phi]= \int\; \sqrt{-g}dt\left[\frac12 \phi^2(t)
+\frac{q^2}{2}\dot{\phi}(t)^2 - \frac14\Phi^4(t) \right],
\label{act}
\end{equation}
where $\Phi=\exp(\frac12 {\cal D}) \phi$, $ {\cal D}=-\pd _t^2-3H(t)\pd _t $
and $H(t)=\dot{a}/a$, $\dot{a}=\partial_t a$.
The Einstein equations have the  form
\bea \label{n1}
3H^2&=&\frac{1}{M^2_p}~~\rho\\
\label{n2}
 H^2+2\ddot{a}/a&=&-\frac{1}{M^2_p}~~p\eea
with the energy and pressure densities are given by \cite{AJK}
\begin{equation}
\label{energy}
 \rho=\frac{q^2}{2} (e^{-\frac12
{\cal D}}\dot{\Phi})^2-\frac{1}{2}
(e^{-\frac12 {\cal D}}\Phi)^2+ \frac14 \Phi^4 +{\cal E}_{1}+{\cal E}_{2}]
\end{equation}
\begin{equation}
\label{pressure} p=\frac{q^2}{2}
(e^{-\frac12 {\cal D}}\dot{\Phi})^2+\frac{1}{2} (e^{-\frac12
{\cal D}}\Phi)^2- \frac14 \Phi^4 -{\cal E}_{1}+{\cal E}_{2}.
\end{equation}
where
\beq
\label{nonloc1}
{\cal E}_{1}=
-\frac{1}{2} \int_0^1 d\rho( e^{\frac{1}{2} \tau {\cal D}}\Phi^3
){\cal D} e^{-\frac{1}{2} \tau{\cal D}} \Phi
\eeq
\beq
\label{nonloc2}
{\cal E}_2= -\frac{1}{2}  \int_0^1 d\tau(
\partial_{t}e^{\frac{1}{2} \tau {\cal D} }\Phi^3 )\partial_t
e^{-\frac{1}{2}\tau{\cal D}} \Phi
\eeq
Equation of motion for the scalar field is
\begin{eqnarray}
\label{Psi-ap}
 \left(q^2{\cal D}+1\right)e^{-{\cal D}}\Phi =\Phi ^3.
\end{eqnarray}

\subsection{Rolling solution in flat space-time}
Taking $H=0$ in (\ref{Psi-ap})
we get the following equation in the flat space
\begin{eqnarray}
\label{phi-ap}
 \left(-q^2\partial_t^2+1\right)e^{\pd^2}\Phi =\Phi ^3.
\end{eqnarray}
This equation contains infinite number of time derivatives, and
 actually  can be written
 in the integral form. It  has been shown
numerically that for $q^2$ small enough there is a solution that
interpolates between non trivial vacua $\Phi(\pm\infty)=\pm 1$
 and  $\Phi(0)=0$ \cite{yar}.
One can get an approximation to this solution
expanding the exponent in (\ref{phi-ap}) in powers of derivatives
and keeping only the second derivatives,
\begin{eqnarray}
\label{Psi-ap-der} \left((1-q^2)\partial_t^2+1\right)\Phi(t) =
\Phi^3(t).
\end{eqnarray}

\begin{figure}[h!]
\centering
\includegraphics[width=2.5cm]{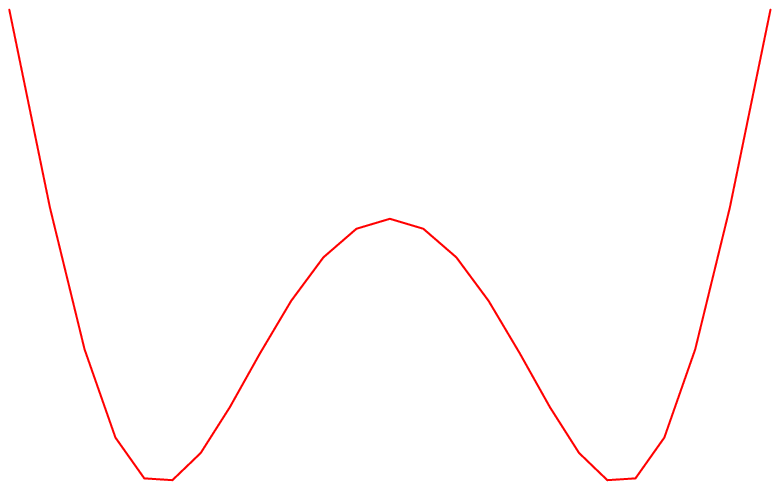}$a)~~
~~~~~ $
\includegraphics[width=2.5cm]{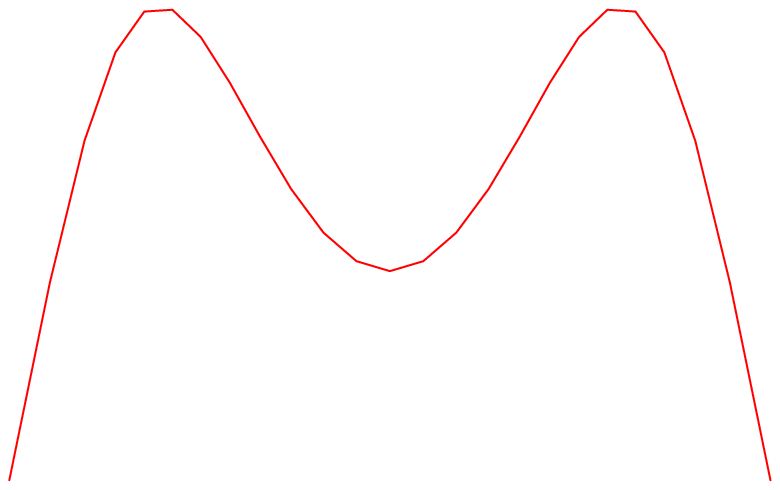}$~~b)$
\caption{Flip of the potential for $q^2<1$ }
 \label{f2}\end{figure}

This  equation describes a particle moving in the potential
$V=\frac{(\Phi^2-1)^2}{4(q^2-1)}+const$. For $q^2<1$ the factor
$q^2-1$ flips the potential, Fig.\ref{f2}.

\begin{figure}[h!]
\centering
\includegraphics[width=3.5cm]{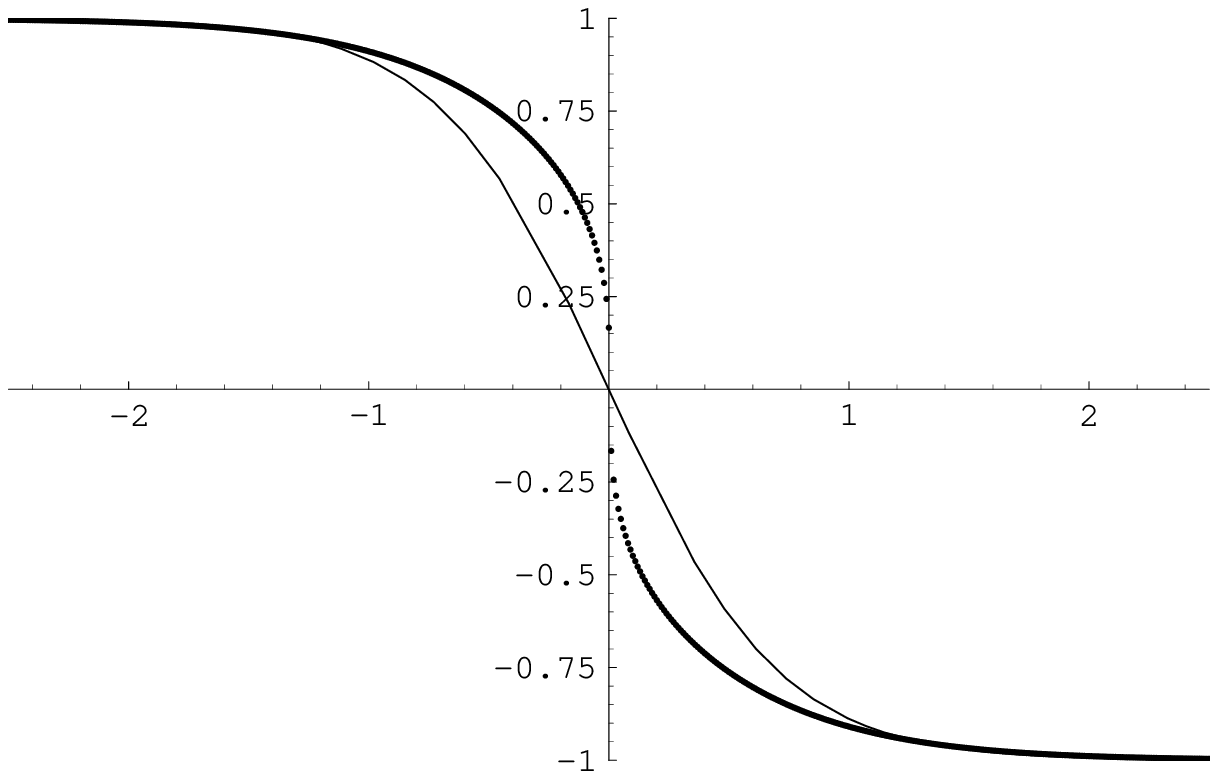}$a~~~$
\includegraphics[width=3.5cm]{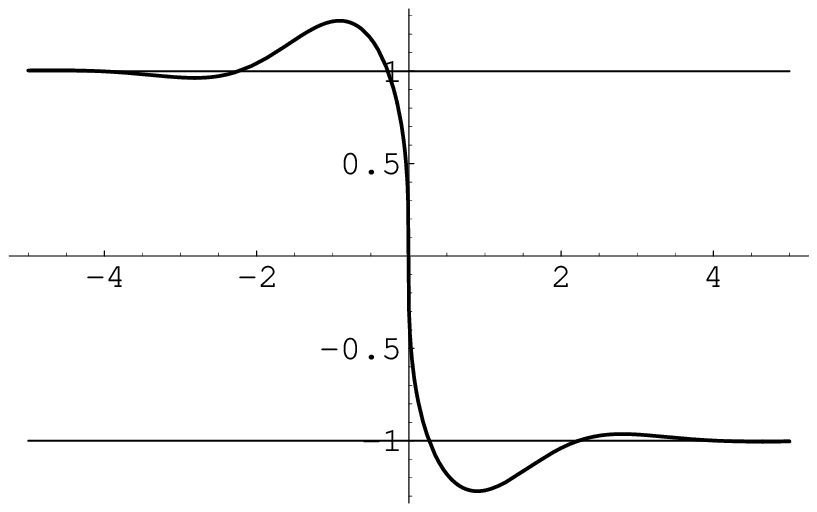}$b~~~~~$
\includegraphics[width=3.5cm]{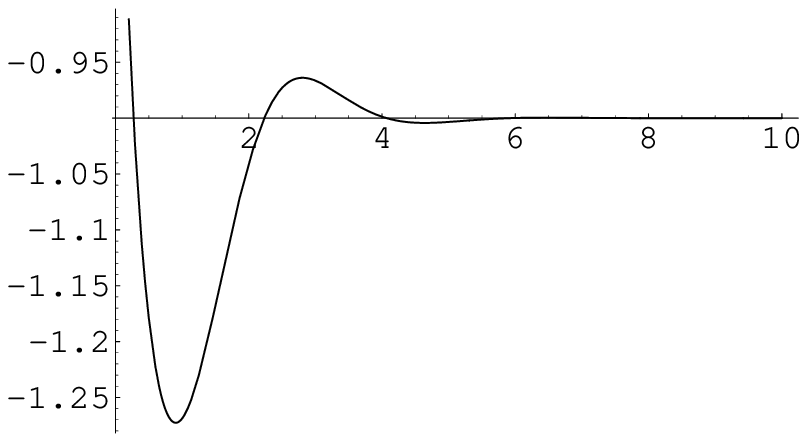}$c$
\caption{a) kink $\Phi_{kink}(t)$ (thin line) and $\Phi_0(t)$ (think line); b)~$\Phi(t)$ for
$q^2=0.96$;$~~~~~$ c) oscillations of $\Phi(t)$ with decreasing
amplitudes around $\Phi_0$}
 \label{f12}
\end{figure}

Equation (\ref{Psi-ap-der})
for $q^2<1$ has
 the  kink solution
$\Phi_{kink}$.
Kink  interpolates between two
vacua during infinitely long time and it is represented in Fig.\ref{f12}a  by a thin line.

 Equation (\ref{phi-ap}) for $q=0$ (the p-adic
string equation of motion for $p=3$) also has a interpolating
solution \cite{q=0,0207107,yar}. We denote it $\Phi _0(t)$ and
plot it in  Fig. \ref{f12}a by think line. Note that the function
$\Phi _0(t)$ is monotonic. From Fig.\ref{f12}a we see that
$\Phi_{kink}$ and $\Phi _0$ have different profiles, but this
difference  is not too big for large times. There is
an essential difference at small time. $\Phi_{kink}$ has the
finite first derivative at $ t=0$, meanwhile the first derivative
of $\Phi _0(t)$ becomes infinite at $t=0$. Note, that the derivative
of the initial scalar field $\phi$ related with $\Phi$ via $\phi=e^{-\frac{1}{2}{\cal D}}\Phi$
is finite at $t=0$. Therefore,
higher derivatives in (\ref{phi-ap}) change the profile of $\Phi
_{kink}(t)$ only at small time and do not change the asymptotic
behavior at large time.

Note, that  small $q^2$    also does not
change too much a profile of a solution  to (\ref{phi-ap}) interpolating between two
vacua. This solution  is plotted  in Fig.\ref{f12}b. The profile of this solution is not a
monotonic function. It can be presented as $\Phi (t) =\Phi_0(t)+\phi(t)$,
where $\phi(t)$
describes  oscillations around $\Phi _0$ with
decreasing amplitude. These oscillations
 are presented in Fig.\ref{f12}c.

\subsection{Approximate solution of system of equations for Non-BPS tachyon
in Friedmann space-time}

Motivated by the flat case we make in (\ref{Psi-ap}) an
approximation
\beq
\exp(\pd _t ^2+3H(t)\pd _t)\Phi \approx (~1+\pd
_t ^2+3H(t)\pd _t~)~\Phi
\eeq
and keep only terms linear on
$(1+\pd _t ^2+3H(t)\pd _t~)$. It is evident that this equation
can be obtained from the action
\begin{equation}
\label{action-M} S^\prime_{scalar}  = \int \sqrt{-g}d^4x\,
\left(\frac{1-q^2}{2} g^{\mu \nu}\pd_\mu \Phi \pd_\nu \Phi
-V(\Phi) \right)\ .
\end{equation}
 We see that  for $q^2<1$ we get
the ghost sign in front of the kinetic terms.
Assuming that $q^2<1$ we take for simplicity  in the following formula $q^2=0$
($q^2<1$ can be achieved  just by rescaling of time).
The corresponding Einstein equations have the form (\ref{energy}), (\ref{pressure})
with
\bea
\label{1'}
\rho &=&-\frac12 \dot{\Phi}^2+V(\Phi),\\
p &=&-\frac12 \dot{\Phi}^2-V(\Phi)
\eea
and the equation for $\Phi$ field
read
\beq
\label{e.o.m.'}
\ddot{\Phi}+3H\dot{\Phi}=V'_{\Phi}
\eeq

Excluding $H$ from (\ref{energy}) and (\ref{e.o.m.'}) one gets the
following equation
\bea \label{BE}
\ddot{\Phi}+\frac{1}{M_p}\dot{\Phi}~\sqrt{3[-\frac12
\dot{\Phi}^2+V(\Phi) ]}-V'_{\Phi}=0
\eea
We see that an equation
similar to the usual  scalar field equation in the Friedmann metric.
There are only  two different signs, one in front of the
kinetic energy in the square root, and the second in front of the
derivative of the potential. It is evident that solutions of this
equation in the slow-roll regime are the same as in the usual case
with "-" potential. However there are differences in the fast roll
regime. The equation state parameter $w$ \bea \label{omega}
w=\frac{p}{\rho}=\frac{\frac12 \dot{\Phi}^2+V(\Phi)}{\frac12
\dot{\Phi}^2-V(\Phi)} \eea is always less then -1, since $w$ can
be represented also  as \bea \label{omega-H}
w=-\frac{3H^2+2\dot{H}}{3H^2}=-1-\frac23 \frac{\dot{H}}{H^2} \eea
and from the equation of motions follows \bea \label{phi-dot-H}
\dot{H} =\frac{1}{2M^2_p}\dot{\Phi}^2, \eea i.e. $\dot{H}$ is
positive.

\subsection{Numerical solutions}
Let us examine numerically solution of the system of equations
(\ref{1'}) and (\ref{e.o.m.'}) for the potential
\bea
\label{pot}
V(\Phi)=\frac{1}4\left(\Phi^2-1\right)^2
\eea

\begin{figure}[!h]
\centering
\includegraphics[width=4cm]{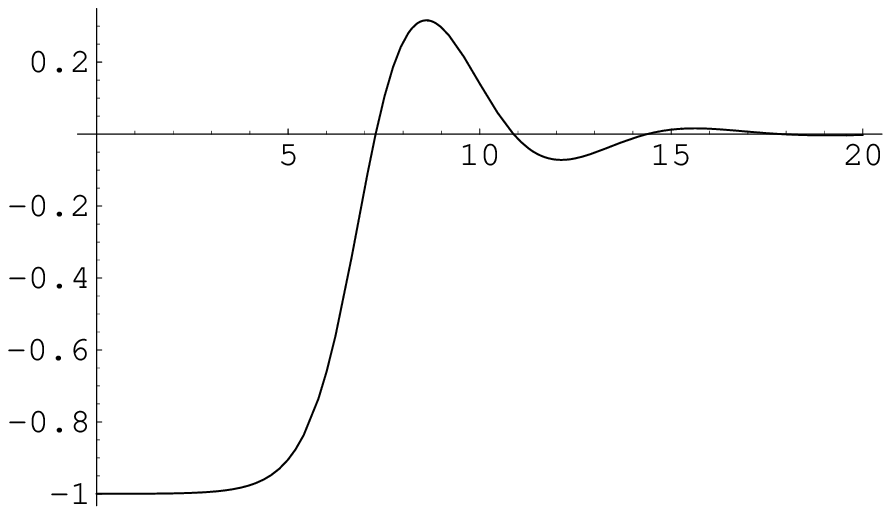}$a~$
\includegraphics[width=4cm]{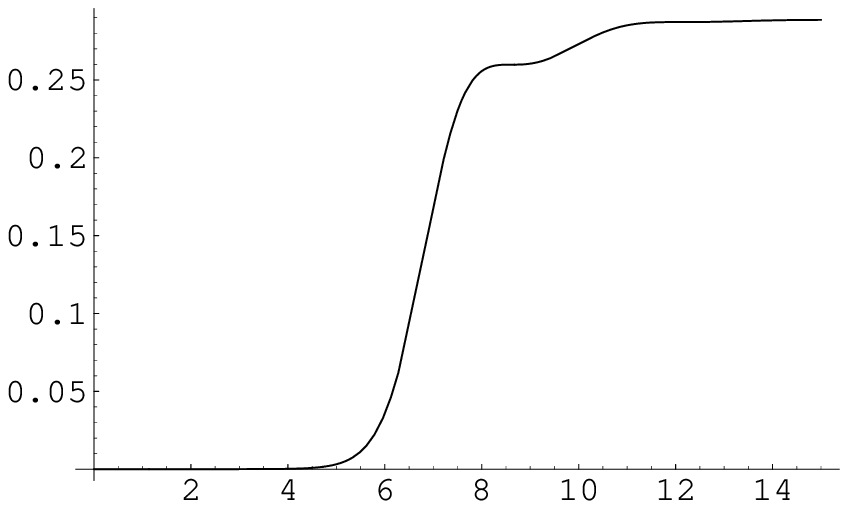}
$b$
\includegraphics[width=3.5cm]{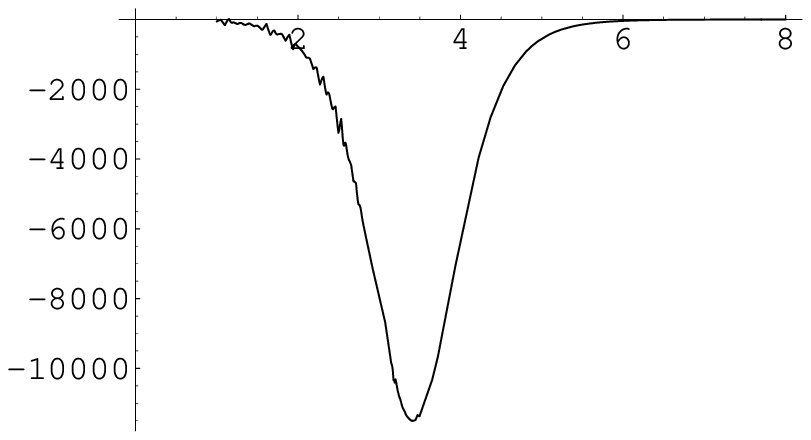}$c$
\includegraphics[width=3.5cm]{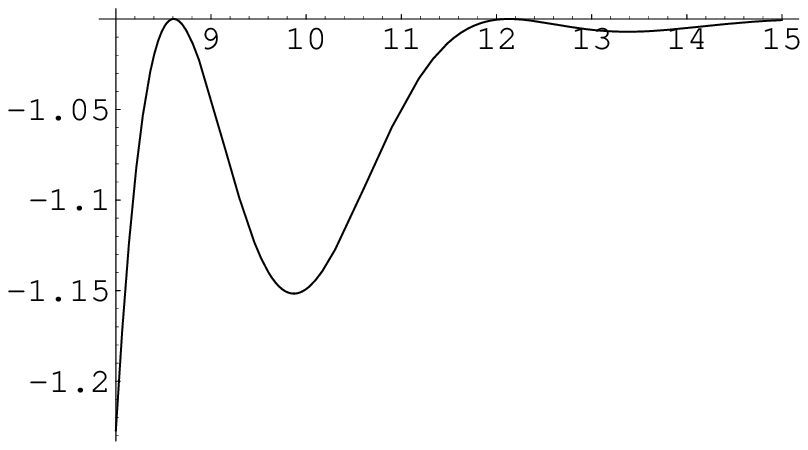}$d$
\caption{ a) Plot of $\Phi=\Phi(t)$ with $\Phi(0)=-1$ and
$\dot{\Phi}\simeq 0$; b) plot of $H=H(t)$; plot of $w=w(t)$
 for
c) $0<t<8$
and  d) for $8<t<15$
}
\label{f-hill-v0}
\end{figure}
There are two  independent initial
conditions for $\Phi(0)$ and  $\dot {\Phi}(0)$. If the initial position $\Phi(0)$
is on the the top of the hill (for the flip potential, Fig.\ref{f2}.b), $\Phi(0)=-1$, and
the initial velocity  is very small  $\dot {\Phi}(0)\simeq 0$
(this corresponds to $H(0)\simeq 0$) then
 after some time   $\Phi$  reaches  the largest position and goes
back to the bottom, and then performs few oscillations and
stops at the
bottom. The final value of $H$ is $1/2\sqrt{3}$. The
evolutions of the scalar field and log-derivative of the scale factor
 are represented in
Fig.\ref{f-hill-v0}.a and Fig.\ref{f-hill-v0}.b. The evolution
of the state equation parameter $w$ is plotted in
Fig.\ref{f-hill-v0}c,d. It starts from -1, becomes a very big
negative number when the field passes  the bottom of the flip
potential Fig.\ref{f-hill-v0}c and goes with small fluctuations
to $-1$ at large times. Fig.\ref{f-hill-v0}.d shows that these
fluctuations do not exceed $-1$.

To reach the top of the hill $\Phi=1$ one has to  increase the velocity, but since
there is a restriction on the initial velocity $\dot
{\Phi}(0)^2\leq 2 V(0)$, (the initial energy should be positive),
 one has to  add a positive constant $V_0$
to the potential to be able to increase the initial velocity.

\begin{figure}[!h]
\centering
\includegraphics[width=5cm]{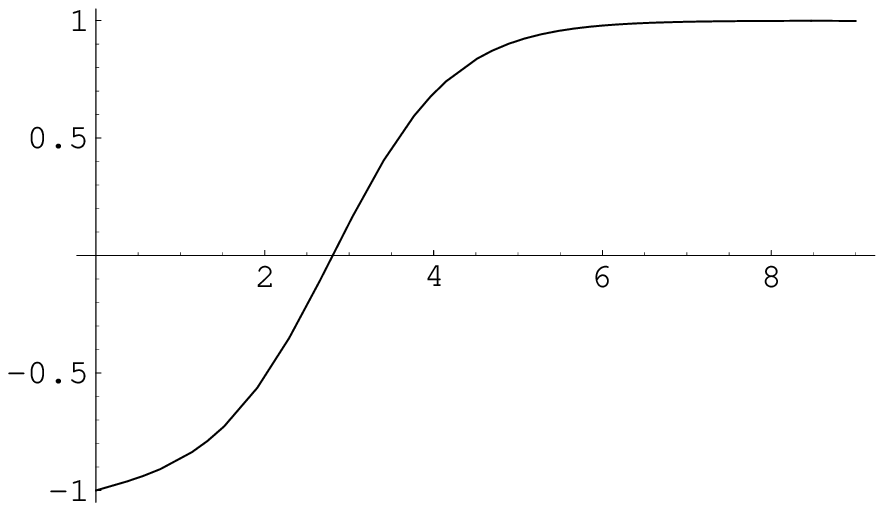}$~~~~~$
\includegraphics[width=5cm]{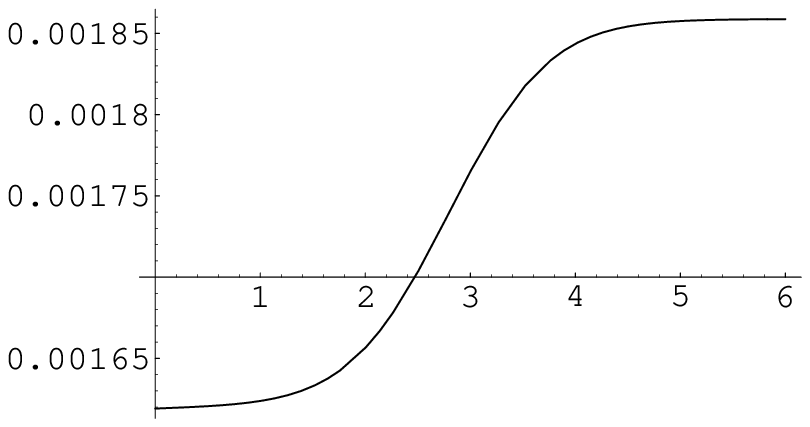}$~~~~~$
\includegraphics[width=5cm]{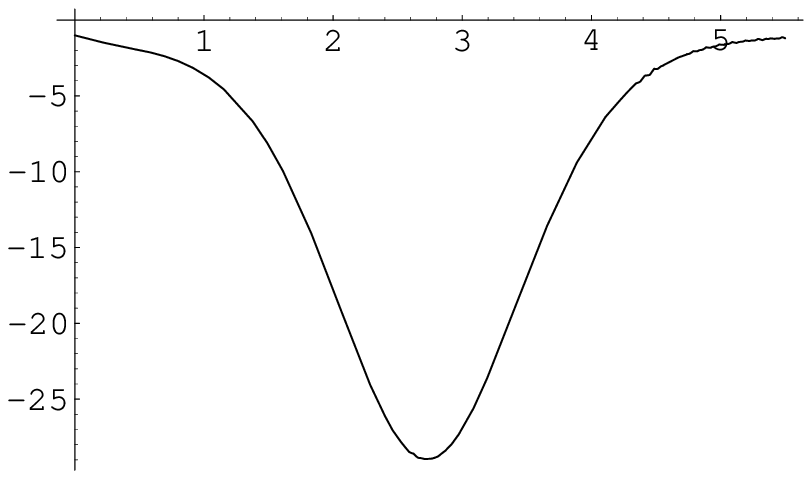}
\caption{Plots of $\phi(t)$, $H(t)$ and $w(t)$ for  $\phi
(0)=-1, \dot {\phi}(0)= 0.1, V_0=0.02$}
\label{f-botom-V}
\end{figure}

For large $M_p$
and  a suitable $V_0$ there is a solution that starts from the  top of one hill with a
non-zero velocity and reach the top of other hill during an infinite time,
Fig.\ref{f-botom-V}. In this case during the initial stages of evolution
the field is near the top of the hill, $\Phi=-1$ and the acceleration is small.
At later times the field begins to evolve more rapidly towards the local minimum
of the flip potential and the equation state parameter $w$ becomes rather big.
Finally, in very late time the field comes closed to the top of other hill,
$\Phi=1$
 \beq
\label{nh} \Phi=1-Ae^{-\alpha t},
\eeq
where $A$ is an arbitrary constant,
$\alpha=(\frac{\sqrt{3V_0}}{M_p}+\sqrt{\frac{3V_0}{M^2_p}+8})/2$
 and a period of $w\simeq -1$, $w<-1$ begins.
This period is infinitely long because the flip potential has the maximum
at $\Phi=1$.

\newpage
\section{Stability}
 In this section we show that fluctuations of the scalar field around the true
vacua are stable for suitable $q^2$, namely we show that the Euclidian action
is positive defined.

 For the usual Goldstone model in the flat space-time
(\ref{pot})
 fluctuations around one of minima $\Phi_0=\pm 1$ correspond to
massive excitations. In our case due to nonlocal factors in the interaction we
get a different picture. Depending on parameter $q^2$ we get two excitations with different
positive square mass or do not get any excitation at all.

Indeed, for fluctuations around $\Phi_0= 1$
\begin{equation}
\label{f-tfl}
\Phi =\Phi_0+\delta \Phi
\end{equation}
we get the quadratic part of the action
\begin{equation}
\label{f-gh}
S_0=\int\delta\Phi\left((q^2\Box
+1)e^{-\Box}-3\right)\delta\Phi dx
\end{equation}

\begin{figure}[!h]
\centering
\includegraphics[width=10cm,angle=-90]{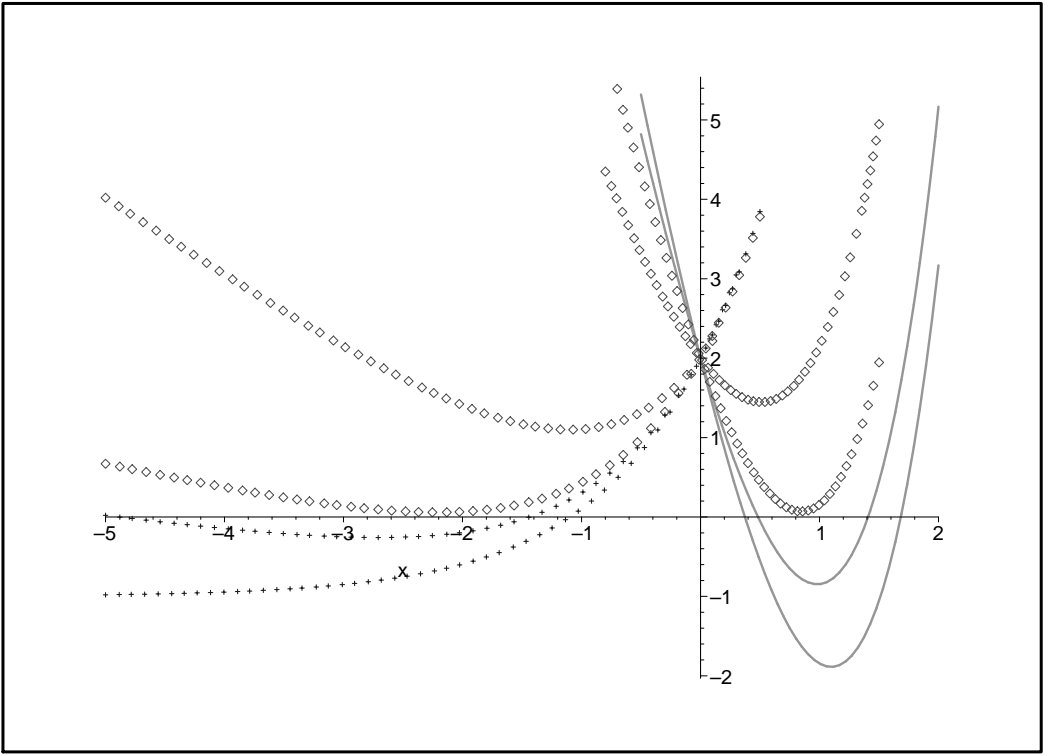}
\caption{Plots of function $f(x)=-q^2x-1+3e^{2x}$
 for  different values of $q^2$$:
q^2<q^2_{cr1}$ (cross lines),
 $q^2_{cr1}<q^2<q^2_{cr2} $ (diamond lines),
$q^2>q^2_{cr2} $ (solid lines)$;
 q^2_{cr1}\simeq 0.33$, $q^2_{cr2}\simeq 7$}
 \label{spectrum}
\end{figure}

To get the particle spectrum we are looking for solutions of the equation
\begin{equation}
\label{f-m-eq}
-q^2 m^2-1+3e^{m^2}=0
\end{equation}
with $=-k_0^2+\vec{k}^2\equiv k^2=-m^2<0$. To find solutions of
this equation we plot in  Fig. \ref{spectrum} the function
\begin{equation}
\label{f-f}
f(x)=-q^2x-1+3e^{x}
\end{equation}
 and find its zeros.

We see that for $q^2<q^2_{cr2}$ this  curve has no positive zeros. For
$q^2>q^2_{cr2}$ there are two zeros, $f(m^2_i)=0$ with
$m_i^2>0,~~~i=1,2$.  Therefore
there are no massive excitations for $q^2<q^2_{cr2}$, and there are
two massive particles for $q^2>q^2_{cr2}$. There is one massive particle
for $q^2_{cr}$.

For $q^2<q_{cr1}$ the function $f(x)$ has also zero for a negative argument
(see "cross-lines" on Fig.\ref{spectrum})
that corresponds to an appearance of tachyons.

Let us note that our propagator around new vacuum is non-standard one. If
we expand $\exp(-\Box)$ in power of derivatives keeping only the first order terms on $\Box$
we get
\begin{equation}
\label{f-gh-k-ap} S_0\approx -\int\delta\Phi(k)\left((q^2-1)k^2
+2\right)\delta\Phi(-k)dk
\end{equation}
and we see that the case $q^2<1$ looks like as if the ghosts are
appearing. In particular, this means that if we performing the
Euclidean
 rotations $k^2=-\omega ^2+\vec{k}\to k^2_E$,
 then our approximated Euclidean propagator $\left((q^2-1)k_E^2
+2\right)$ for $q^2<1$
 it is not positively defined.
However, the full propagator that one gets from  $S_0$ after
the Euclidean
 rotations
\begin{equation}
\label{f-gh-kE} S_0=-\int\delta\Phi(k)\left((q^2k_E^2
-1)e^{k_E^2}+3\right)\delta\Phi(-k)dk
\end{equation}
is positive defined for $q^2_{cr1}<q^2<q^2_{cr2} $.
Indeed,
\begin{equation}
\label{f-gh-kE}
\left((q^2k_E^2
-1)e^{k_E^2}+3\right)=f(-k_E^2)e^{k_E^2}
\end{equation}
and  we see form Fig.\ref{spectrum} that the  function f(x)  for negative
arguments is positive for all negative arguments (positive
$k^2_E$) if $q^2>q^2_{cr1}$. Therefore, we see that the inverse
propagator is strictly positive for $q_{cr2}^{2}>q^2>q_{cr1}^{2}$.
It is interesting to note that for $q^2=0$ the inverse propagator
$f(-k_E^2)e^{2k_E^2}$ is not positive.

\section{Conclusion}
We have studied  the evolution of open GSO $-$ NSR string tachyon
in the Friedmann space-time. The corresponding  solution in the flat space-time is
known as a rolling tachyon and it describes the decay of the space
filling D3 brane corresponding to the unstable perturbative vacuum
to the local stable vacuum. We have performed calculations under the
following approximations and assumptions:
\begin{itemize}
\item
the level truncation and an approximation of a slow
varying axillary field;
\item  a direct generalization of
the  tachyon nonlocal action to the Friedmann space-time;
\item
an effective local action approximation.
\end{itemize}

We have found that in the
effective field theory approximation the equation of state
parameter $w<-1$, i.e. one has a phantom Universe,
but
there is no problem with  quantum instability without this approximation.
We have found  that to reach the nonperturbative vacuum
one has to add to the action a brane tension which is
larger that is required by the Sen hypothesis.
This large  brane tension can be interpreted as an
effect of the closed string excitations.

It would be interesting to try to find an analog
 of this solution to a full SFT theory (without level truncation),
 in particular with a framework of
 Vacuum String Field Theory. VSFT is a version of usual SFT that is supposed
 to describe the theory
 at the minimum
 of the tachyon potential. Corresponding solutions for bosonic VSFT
 have been found recently
 in  \cite{LB}. It would be also interesting to see
if one has to shift
 the tension of D-brane to get a rolling solution
in the Friedmann metric.

$$~$$

{\bf Acknowledgements}

I would like to thank L. Bonora, B. Dragovich, A. Koshelev, S. Mukhanov, S. Vernov and I. Volovich
for useful discussions. This work is supported in part by RFBR grant
02-01-00695 and INTAS grant 03-51-6346.
$$~$$

\end {document}